%% file: iclr2025_conference.tex
\documentclass{article} 
\usepackage{iclr2025_conference,times}
\usepackage{graphicx}
\usepackage{hyperref}
\usepackage{amsmath}
\usepackage{algorithm}
\usepackage{algpseudocode}
\usepackage{url}
\usepackage[font=small,skip=0pt]{caption}

\input{math_commands.tex}

\title{Data Curation for Machine Learning \\ Interatomic Potentials by \\ Determinantal Point Processes}

\author{Joanna Zou 
\\
Computational Science \& Engineering\\
Massachusetts Institute of Technology\\
Cambridge, MA, 02139, USA \\
\texttt{\{jjzou\}@mit.edu} \\
\And
Youssef Marzouk \\
Computational Science \& Engineering \\
Massachusetts Institute of Technology \\
\texttt{\{ymarz\}@mit.edu} 
}

\iclrfinalcopy 
\begin{document}

\maketitle

\begin{abstract}
The development of machine learning interatomic potentials faces a critical computational bottleneck with the generation and labeling of useful training datasets. We present a novel application of determinantal point processes (DPPs) to the task of selecting informative subsets of atomic configurations to label with reference energies and forces from costly quantum mechanical methods. Through experiments with hafnium oxide data, we show that DPPs are competitive with existing approaches to constructing compact but diverse training sets by utilizing kernels of molecular descriptors, leading to improved accuracy and robustness in machine learning representations of molecular systems. Our work identifies promising directions to employ DPPs for unsupervised training data curation with heterogeneous or multimodal data, or in online active learning schemes for iterative data augmentation during molecular dynamics simulation. 

\end{abstract}

\section{Introduction}
\label{sec:intro}

A primary challenge in the development of machine learning interatomic potentials (MLIPs) for atomistic simulation is the high degree of sensitivity of model performance to the choice of training set. In practice, parameters of MLIPs are learned from a training set of atomic configurations labeled with reference ground state energy and force values obtained from higher-order quantum mechanical (QM) calculations such as density functional theory (DFT). Due to the significant cost of QM calculations, training datasets must be limited in size to keep the data generation task computationally tractable and reduce overfitting to redundant data, but also retain representation of conformational diversity in order to produce robust MLIPs capable of capturing chemical processes of interest.

In this work, we propose using \textit{determinantal point processes} (DPPs) for automated training set construction for machine learning-driven atomistic simulation. A DPP is an efficient probabilistic model over subsets of discrete sets which assigns greater likelihood to subsets with diverse elements, as determined by the determinant of a kernel matrix measuring the similarity between elements. Our work is one of the first to compare state-of-the-art data subselection algorithms on the task of training MLIPs, assessed in terms of diversity of atomic environments sampled, accuracy of the MLIP as it varies with training set size, and transferrability of the MLIP to predicting energies of atomic environments unseen during training. 

\section{Background and Related Work}
\label{sec:relatedwork}

Data subselection is one form of \textit{active learning}, in which an algorithm queries informative samples from a large pool of unlabeled data to label using a cost-intensive process for regression tasks. For MLIP training, the initial pool of unlabeled data is generally sourced from 1) hand-crafted datasets using expert judgment for each system of study, such as those of the QM9 database \citep{Ramakrishnan2014}; or 2) from time steps of molecular dynamics (MD) simulation which are efficiently evaluated using an empirical potential or initial iterate of the MLIP to approximately sample from the Gibbs distribution of the system. However, since the step size must be sufficiently small for stable numerical simulation (on the order of $10^{-12}$ or $10^{-15}$ seconds), configurations generated by MD simulation are highly correlated and lead to high redundancy in the dataset.

Data subselection techniques seek to assemble compact training sets which improve on hand-crafted datasets as well as the naive approach of uniform sampling from the MD trajectory. In \cite{Huan2017}, the descriptor space is partitioned into clusters using Euclidean distances via $k$-means which are each subsampled randomly. However, the $k$-means algorithm is sensitive to the geometry of the descriptor space and performs poorly in the high dimensional regime, which is often the setting for molecular descriptors. In \cite{Podryabinkin2017, Lysogorskiy2023}, the MaxVol algorithm identifies an optimal subset of configurations whose descriptors span the largest volume, based on the D-optimality criterion in linear experimental design. This approach is limited to choosing a fixed number of data, namely the same number as the dimension of the descriptors, to form a square D-optimal matrix. In the entropy-maximization method in \cite{Karabin2020,Zapiain2022}, local maxima of a chosen entropy function are assumed to be sufficiently non-redundant and included in the training dataset. This method heavily relies on the assumption that the full support of an effective potential function can be adequately sampled such that the local maxima correspond to well-separated modes of the entropy function. Each method utilizes a different similarity metric to compare atomic configurations, summarized in Table \ref{tab:litreview} in Appendix \ref{app:lit}, with the aim to reduce redundancy in the training set.

In the absence of labels, samples must be selected in an unsupervised manner utilizing only information on the atomic environment, which is summarized using \textit{descriptors} which characterize multi-body interactions while preserving invariance properties of the representation. Examples of such descriptors include symmetric polynomials (ACE potential, \cite{Drautz2019}), bispectrum components (SNAP potential, \cite{Thompson2015}), SOAP descriptors (GAP potential, \cite{Bartok2010}), atom-centered symmetry functions \citep{Behler2011}, SMILES strings used in graph neural network (NN) potentials \citep{Weininger1988}, or other latent representations learned simultaneously by NN potentials such as NequIP \citep{Batzner2022} and Allegro \citep{Bartók2022}. In principle, the choice of descriptors for active learning may be independent from the choice of MLIP architecture which is trained, though the expressiveness of the descriptor -- its ability to distinguish between atomic environments -- will affect the efficiency of the data subselection algorithm.

\begin{figure}[hbt!]
    \centering
    \includegraphics[width=1.05\textwidth]{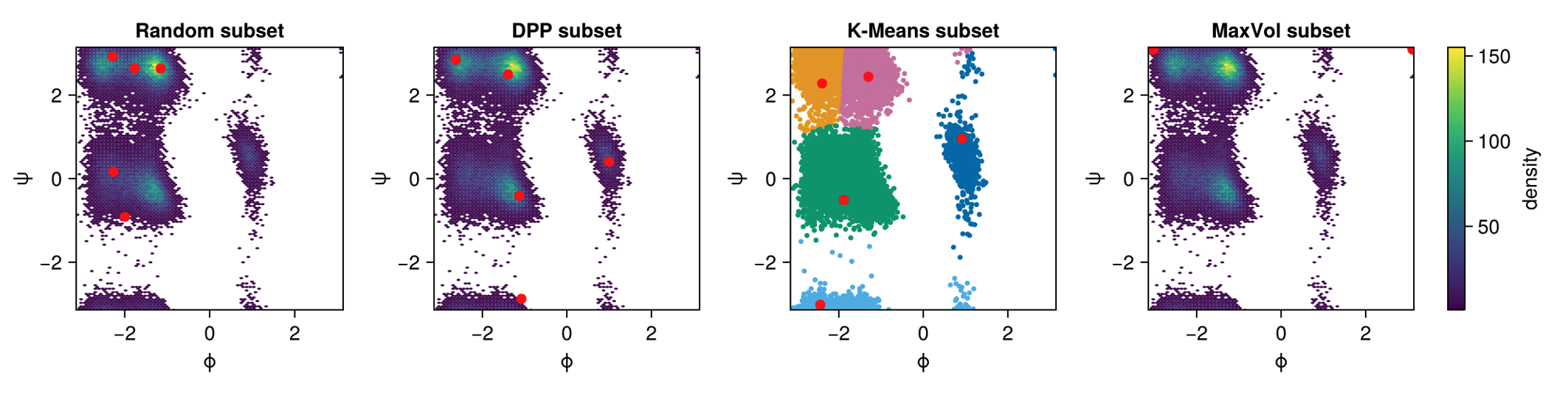}
    \caption{Sample subsets drawn with each method from a MD trajectory of alanine dipeptide.}
    \label{fig:dalanine}
\end{figure}
To illustrate the data subselection methods, Figure \ref{fig:dalanine} shows subsets of $N=5$ configurations of alanine dipeptide, taken from a 50-ns Langevin dynamics simulation at 300 $K$ using the AMBER ff-99SB-ILDN force field, sourced from \cite{markovmodel}. The subsets are drawn using two collective variables (backbone dihedral angles $\phi, \psi$) as the descriptors of the atomic configurations. Unlike the uniform random subset, the subsets drawn with the DPP and $k$-means more consistently distribute the five selected points to distinct regions of descriptor space, utilizing similarity-based metrics and distance-based clustering respectively. The MaxVol algorithm is limited to selecting 2 points (equal to the collective variable dimension), and these configurations are located at the bounds of the descriptor space in order to maximize the volume enclosed by their span.

\section{Methodology}
\label{sec:dpps}

DPPs have seen a recent surge of interest in applications to machine learning \citep{Kulesza2012}, experimental design \citep{Derezinski2020}, and dimensionality reduction \citep{Belhadji2020}. This section gives an intuitive introduction to DPPs, with details in Appendix \ref{app:dpps}. If the data pool is modeled by a point process, where each configuration constitutes a point in descriptor space, then a DPP is a type of repulsive point process which favors the sampling of distinctly unique and non-clustered points. Unlike other repulsive point processes, DPPs have several features, including closed-form probabilities and efficient sampling algorithms, which make them attractive in ML applications \citep{Lavancier2015}. Along this vein, simple random sampling can be interpreted as a homogeneous Poisson process where points are purely independent and uniformly distributed. 

As a point process model, a DPP places a random measure over all subsets of a discrete dataset $\mathcal{Y}$ of $M$ elements. The degree of repulsion between elements is controlled by a kernel matrix $K \in \mathbb{R}^{M \times M}$, where $K_{ij} = \kappa(Y_i, Y_j)$ for some positive semidefinite (PSD) kernel function $\kappa: \mathcal{Y} \times \mathcal{Y} \to \mathbb{R}$. The DPP places greater probability on subsets which have a higher degree of repulsion corresponding to a larger determinant of a kernel matrix, which can be interpreted as a measure of volume spanned by the matrix columns. For a subset $\{Y_i\}_{i \in \bm{\alpha}} \subseteq \mathcal{Y}$ indexed by $\bm{\alpha} = \{\alpha_1, ..., \alpha_m \}$ with $m \leq M$ and the kernel matrix $[K_{ij}]_{ i,j \in \bm{\alpha} }$ restricted to rows and columns indexed by $\bm{\alpha}$, the probability of the subset being drawn by the DPP is given by: 
\begin{equation}
    \label{eq:dpp}
    \mathcal{P}(\{Y_i\}_{i \in \bm{\alpha}}) = \det \big( [K_{ij}]_{ i,j \in \bm{\alpha} } \big)
\end{equation}
A DPP is defined by a choice of PSD kernel function. In this work, we employ a linear kernel equivalent to the cosine similarity of the descriptor vectors. Given a configuration of $J_i$ atoms with Cartesian positions $\mathbf{x}_i \in \mathbb{R}^{3J_i}$, a second configuration of $J_j$ atoms with positions $\mathbf{x}_j \in \mathbb{R}^{3J_j}$, and a global $q$-dimensional descriptor $\phi: \mathbb{R}^{3J} \to \mathbb{R}^q$ as a function of the positions of an arbitrary number of atoms $J$, the normalized linear kernel is given by: 
\begin{equation}
    \label{eq:dpkernel}
    \kappa(\mathbf{x}_i, \mathbf{x}_j) = \frac{\phi(\mathbf{x}_i) \cdot \phi(\mathbf{x}_j)}{ \lVert \phi(\mathbf{x}_i) \rVert \ \lVert \phi(\mathbf{x}_j) \rVert}
\end{equation}
A DPP models probabilities over all $2^M$ subsets of $\mathcal{Y}$ without constraints on the size of the subset; however, subsets of a pre-specified size are often desired in applications. Therefore, we employ fixed-size DPPs \citep{Kulesza2011, Barthelme2019} which provide a probability measure over subsets of $\mathcal{Y}$ of the same cardinality. Efficient sampling algorithms have been developed for both regular and fixed-size DPPs which perform sampling by decomposing the DPP into a mixture of elementary DPPs; refer to \cite{Kulesza2011, Barthelme2019} and Appendix \ref{app:dpps}. In general, the probabilistic nature of DPPs can offer substantially improved computational efficiency over brute force or optimization-based methods to data subselection.

\section{Experiment}
\label{sec:experiment}
\textbf{Reference dataset.} We utilize hafnium oxides as systems of study for benchmarking the performance of the data subselection algorithms. Training and validation data belong to a 45,201-element set of HfO\textsubscript{2} and HfO configurations generated from six simulation sets to represent a range of states across the compression curve of each system, starting from crystal structures obtained from the Materials Project database \citep{Jain2013}. A 67\% partition of this set, chosen proportionally from each simulation setting, is taken to be candidate data from which training data are subselected, while the remaining 33\% constitutes the validation set. The test data consist of 67,219 configurations of either hafnium (Hf) or oxygen (O) atoms generated across 12 simulation sets by a similar procedure. Using this test set, we evaluate the ability of the MLIP trained on multi-species configurations of hafnium oxides to extrapolate to single-species configurations of Hf\textsubscript{2}, Hf, O\textsubscript{2}, and O, which are not explicitly seen during training. Reference QM values of energies and forces are computed with DFT using Quantum ESPRESSO \citep{Giannozzi2017}. Further details are provided in Appendix \ref{app:refdata}. 

\textbf{Experimental setup.} Data subselection is performed with four methods, by taking 1) a uniform sample of the candidate data, referred to as a simple random sample (SRS); 2) a random sample from clusters identified with $k$-means using $k=5$ and covariance-weighted Euclidean distances between descriptors, where clusters are uniformly sampled proportionally to the cluster size; 3) the fixed set of basis vectors selected by the MaxVol algorithm; and 4) a random sample from a fixed-sized DPP computed from a linear kernel matrix of the descriptors. The molecular descriptors used to represent configurations are global ACE descriptors of Hf-O systems with body order $N=5$ and polynomial degree $p=8$ \citep{Drautz2019}, leading to descriptors of dimension $d=1160$. Once training data are selected, reference QM values of energy are queried and an ACE potential is trained on the set by taking the least squares estimator of the model parameters.

\begin{figure}[hbt!]
    \centering
    \includegraphics[width=0.8\textwidth]{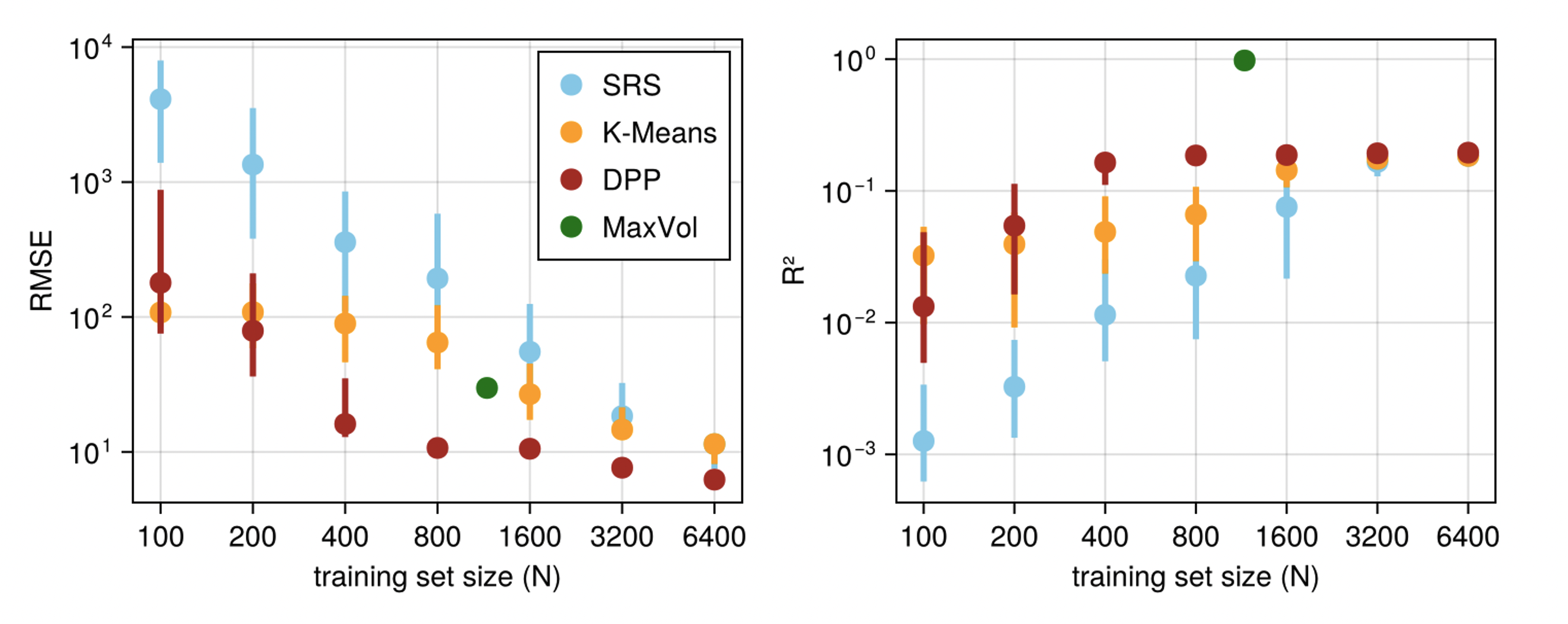}
    \caption{RMSE and $R^2$ values of energy predictions on the validation set by the MLIP trained on set size $N$. The center point denotes the median value and rangebars denote the 25th to 75th percentile over 100 trials.}
    \label{fig:train_accuracy}
\end{figure}

\textbf{Accuracy vs. training set size.} We compare the accuracy of a MLIP trained on data subselected with each method using two common accuracy metrics: root mean square error (RMSE) in energy predictions capturing bias and variance information, and the coefficient of determination ($R^2$) measuring goodness of fit. For algorithms which sample variable set sizes (SRS, $k$-means, and DPP), accuracy statistics over 100 trials on the validation data are reported in Figure \ref{fig:train_accuracy} for set sizes $N \in \{100, 200, 400, 800, 1600, 3200, 6400\}$, whereas the accuracy is reported for a single set size $d=1160$ for MaxVol, as by construction it solves for a deterministic subset with cardinality equal to the dimension of the descriptor vector. The DPP-trained MLIP is comparable to that trained with $k$-means in both the data-poor regime ($<$200 samples), where small training sets limit model accuracy, as well as in the data-rich regime ($>$6400 samples), where large training sets have reduced distinction from one another. DPP outperforms the other variable-size data subselection methods, achieving low RMSE and high $R^2$ for each training set size $N$ with less dispersion as represented by the interquartile range. MaxVol achieves an $R^2$ value closest to 1. This study demonstrates that DPP-based subselection is effective at constructing training datasets which summarize the Hf-O data with $\mathcal{O}(10^2-10^3)$ elements. 

Although SRS, $k$-means, and DPP are flexible to select variable set sizes, we fix the set size sampled by each of these methods to $N = d=1160$ for the remainder of the studies, in order to maintain one-to-one comparison with the MaxVol algorithm. 

\begin{figure}[hbt!]
    \centering
    \includegraphics[width=0.72\textwidth]{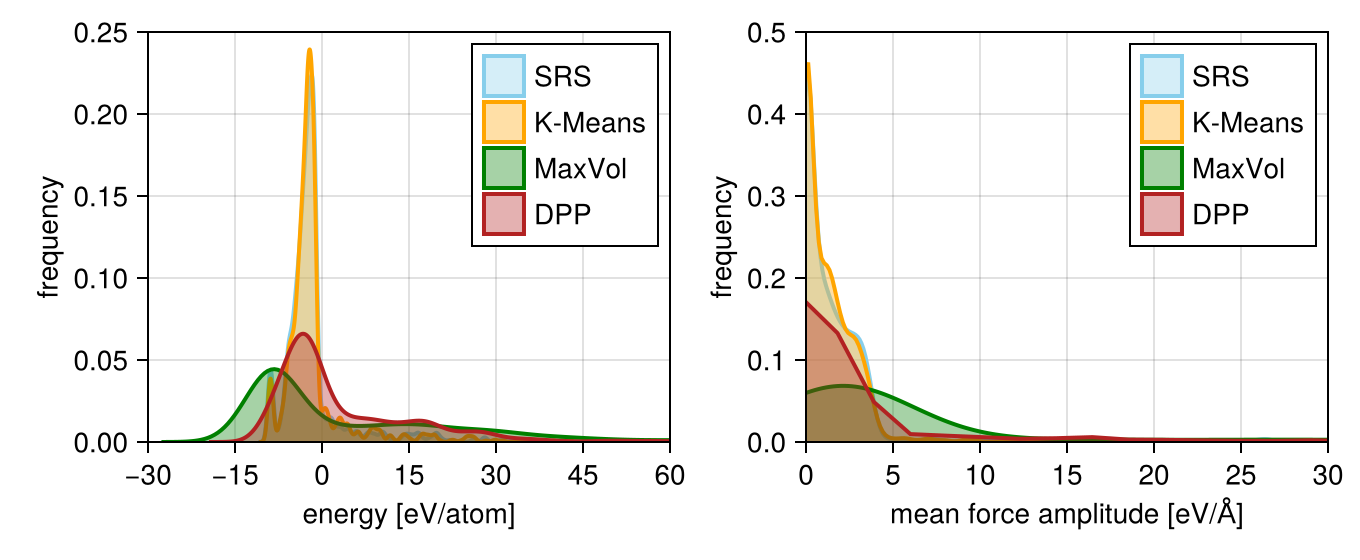}
    \caption{Distribution of energies and force amplitudes of 1160 selected configurations.}
    \label{fig:ef_rep}
\end{figure}

\textbf{Diversity of training set.} The subsets chosen by each method is assessed for diversity, which is measured in terms of reference energies and force amplitudes (averaged over all atoms) associated with configurations in the subset, following \cite{Zapiain2022}. In Figure \ref{fig:ef_rep}, the subsets drawn with DPP and MaxVol cover a greater range of energies and force amplitudes, whereas the subset drawn with $k$-means has marginal distinction from that drawn with SRS, indicating that distance-based clustering in descriptor space does not necessarily correspond to better coverage of output quantities. This study provides evidence of the expressiveness of product-based similarity metrics used by DPPs and MaxVol to choose subsets with a greater range in the output space.

\begin{figure}[hbt!]
    \centering
    \includegraphics[width=0.9\textwidth]{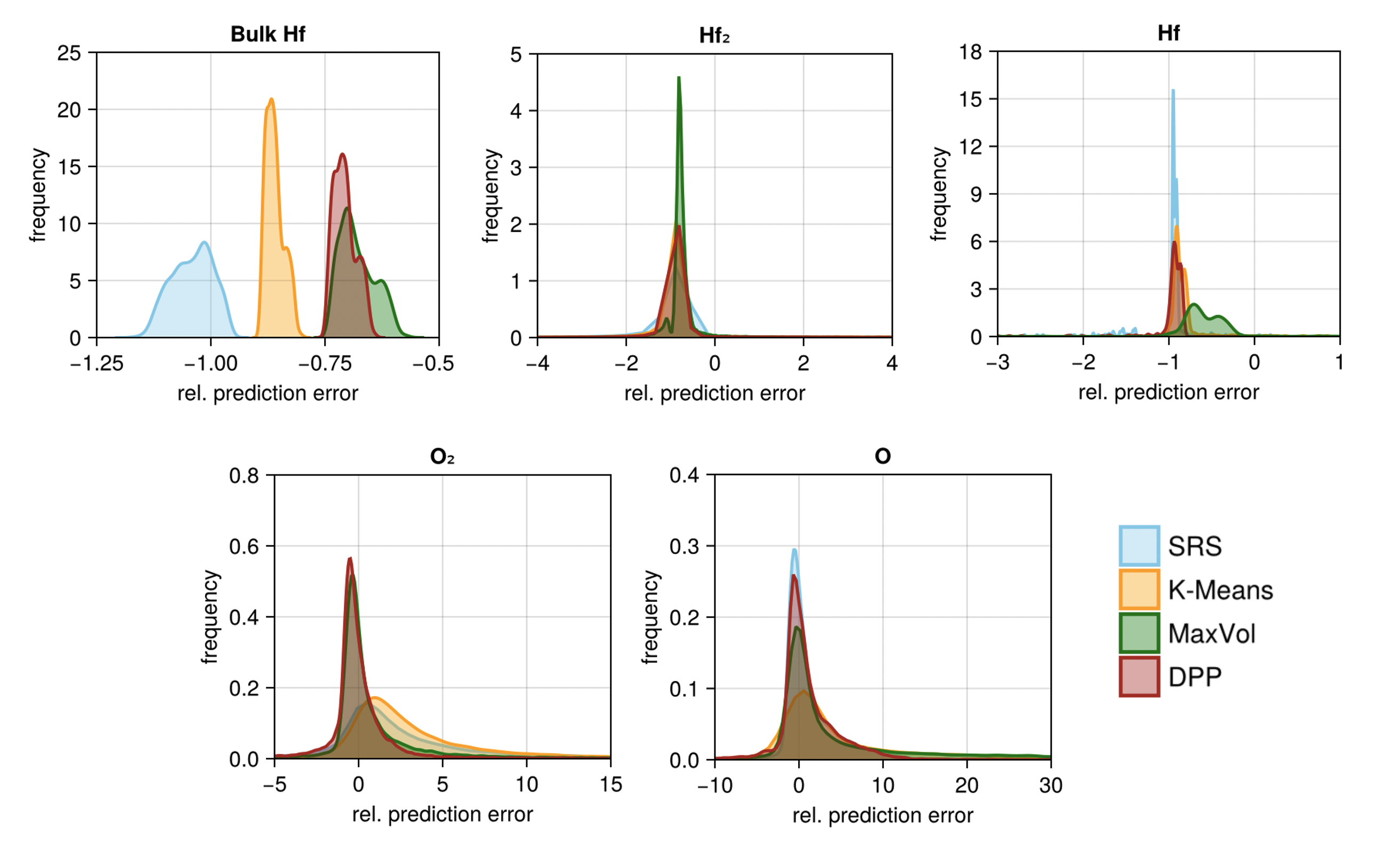}
    \caption{Relative error in energy predictions on the test set by an MLIP trained on a 1160-element set. Errors in Hf\textsubscript{2}, Hf, O\textsubscript{2}, and O have been truncated to visualize the main mass of the distribution.}
    \label{fig:prediction}
\end{figure}

\textbf{Prediction error on unseen data.} We evaluate the generalization capabilities of the MLIP, trained on atomic interactions with HfO\textsubscript{2} and HfO data, to predict energies of single-species Hf and O systems. Figure \ref{fig:prediction} shows the relative error in energy predictions over test configurations categorized into 5 molecule types, as summarized in Appendix \ref{app:refdata}. In general, the prediction error by the DPP-trained MLIP is close to that of the MaxVol-trained MLIP, which achieves the lowest distribution of error for the bulk 128-atom Hf system.

\section{Discussion}
\label{sec:conclusion}

We present a new approach to training data curation for MLIPs leveraging DPPs and provide one of the first studies to benchmark the performance of different data subselection algorithms, advancing the state of the art in computational materials characterization. Our work demonstrates the competitiveness of the DPP-based approach with respect to existing approaches to variable-size data subselection in terms of accuracy of the trained MLIP, diversity of sampled configurations, and generalization to predictions on unseen data. If the kernel matrix is chosen to be the Fisher information matrix, then DPPs can be viewed as a principled probabilistic counterpart to the MaxVol algorithm, as both approaches rely on the D-optimality principle of maximizing the matrix determinant \citep{Derezinski2020}. This technique may complement several existing efforts to accelerate active learning in an online setting, where data are simultaneously sampled via molecular dynamics simulation of a fast potential and assessed for labeling with calls to a costly QM method. The next step of this work is to employ conditional DPP sampling for data augmentation tasks amenable to online active learning, to draw subsets of novel configurations conditioned on an existing set of training configurations, and compare this kernel-based strategy to uncertainty-based strategies proposed in \cite{Vandermause2020, vanderOord2023, Kulichenko2023}. Another future direction is to apply this method to heterogeneous or multimodal datasets, such as simulation data of multiple element species or combinations of simulation and experimental data, as an automated technique for extracting informative data subsets which are most efficient for model development.

\subsubsection*{Acknowledgments}
The authors would like to thank Dallas Foster for discussions on this work and Dionysios Sema for supplying the reference dataset of DFT calculations. JZ and YM gratefully acknowledge support from the United States Department of Energy, National Nuclear Security Administration under Award Number DE-NA0003965.

\bibliography{iclr2025_conference}
\bibliographystyle{iclr2025_conference}

\include{appendix}

\end{document}

%% file: math_commands.tex
\usepackage{amsmath,amsfonts,bm}

\def\eqref#1{equation~\ref{#1}}

\def\1{\bm{1}}

\DeclareMathAlphabet{\mathsfit}{\encodingdefault}{\sfdefault}{m}{sl}
\SetMathAlphabet{\mathsfit}{bold}{\encodingdefault}{\sfdefault}{bx}{n}



%% file: appendix.tex
\appendix
\section{Appendix}
\label{sec:app}

\subsection{Comparison of data subsampling algorithms}
\label{app:lit}
Table \ref{tab:litreview} provides a summary of the literature review from Section \ref{sec:relatedwork}. A primary distinction among the algorithms for data subselection is the similarity metric used to compare two configurations. While the entropy maximization method is amenable to offline active learning schemes for data subselection, the entropy function depends on local (per-atom) descriptors, and is therefore excluded from comparison studies of methods depending on global (molecular) descriptors.

\begin{table}[hbt!]
\fontsize{8}{7}\selectfont
\caption{Review of data subselection algorithms for ML-IPs in literature.}
\begin{tabular}{ccccc}
\hline
\textbf{Algorithm}     & \textbf{Primary Reference}     & \textbf{Similarity Metric}             & \textbf{Subset Size} & \textbf{Type} \\ \hline
SRS & --                             & None                                   & Variable             & Probabilistic                  \\
$k$-means clustering     & \cite{Huan2017}             & Euclidean distance in descriptor space & Variable             & Probabilistic                  \\
MaxVol                 & \cite{Podryabinkin2017} & Fisher information matrix                            & Fixed                & Deterministic                  \\
Entropy max.   & \cite{Karabin2020}        & Entropy function                       & Variable             & Deterministic                  \\
fixed-size DPP                  & This work                      & PSD kernel                             & Variable             & Probabilistic                  \\ \hline
\end{tabular}
\label{tab:litreview}
\end{table}

\subsection{Reference dataset}
\label{app:refdata}

The reference data used to validate energy predictions in the experiments of Section \ref{sec:experiment} consist of quantum-level energies and forces of atomistic configurations of hafnium (Hf), oxygen (O), hafnium dioxide (HfO\textsubscript{2}), and hafnium oxide (HfO) systems. The configurations are generated using a systematic approach to explore the descriptor space of interest: for each system, the experimentally observed crystal structures are obtained from Materials Project \citep{Jain2013}. Using the compression curves of each material, Latin Hypercube Sampling (LHS) is performed, treating the lattice parameters (length and angles) and density as random variables, to draw random samples ranging from highly compressed states to melting point states. The atomic positions are then minimized for each configuration to obtain the equation of state (EOS). Additional simulations are carried out for bulk Hf systems to sample configurations along the phase diagram from the convex hull to high energy states In particular, a modified Monte Carlo rattling procedure using the hiPhive package \citep{Eriksson2019} is employed on supercells replicated from the unit cell of each structure to generate LHS samples emulating the phase diagram. For each configuration, quantum-level energies are computed with Quantum ESPRESSO \citep{Giannozzi2017} utilizing the PBE functional with the Optimized Norm Conserving Vanderbilt (ONCV) pseudopotential. All calculations used the Marzari-Vanderbilt method \citep{Marzari1999} for electron smearing and the electronic temperature was set to 0.06 Ry. The kinetic energy cutoff was 90 Ry and the k-point spacing of 0.025 Å\textsuperscript{-1} was adopted to ensure all configurations have consistent spacing.

Table \ref{tab:refdata} summarizes the configurations of HfO\textsubscript{2}, HfO used for training and validation and Table \ref{tab:testdata} summarizes the configurations of Hf and O systems used for the test set, categorized into simulation sets. The simulation sets are labeled with their chemical species; Materials Project ID number (MP ID), if one exists; the number of translation and/or rotational degrees of freedom (1D, 3D, 6D); phase (primitive, gas); whether the Monte Carlo rattling procedure is performed (MC); and whether the atomic positions have been minimized (EOS). Set 1 of Table \ref{tab:refdata} (labeled with ``figshare'') corresponds to hafnium dioxide configurations sourced from \cite{Sivaraman2020} where electronic structure calculations are recomputed with the ONCV pseudopotential. The number of Hf or O atoms per configuration are reported in the last two columns.

\begin{table}[hbt!]
\caption{Training/validation set of hafnium dioxide (HfO\textsubscript{2}) and hafnium monoxide (HfO) configurations.}
\centering
\begin{tabular}{llllll}
\hline
\textbf{index} & \textbf{MP ID} & \textbf{set} & \textbf{no. configurations} & \textbf{no. Hf} & \textbf{no. O} \\ \hline
1              & --             & HfO\textsubscript{2} figshare    & 2052                        & 36              & 72             \\
2              & 352            & HfO\textsubscript{2} EOS 1D      & 300                         & 4               & 8              \\
3              & 550893         & HfO\textsubscript{2} EOS 1D      & 131                         & 1               & 2              \\
4              & 550893         & HfO\textsubscript{2} EOS 6D      & 27620                       & 1               & 2              \\
5              & --             & HfO gas          & 129                         & 1               & 1              \\
6              & --             & HfO EOS 1D       & 14969                       & 1               & 1              \\ \hline
\end{tabular}
\label{tab:refdata}
\end{table}

\begin{table}[hbt!]
\caption{Test set of hafnium (Hf) and oxygen (O) configurations.}
\centering
\begin{tabular}{llllll}
\hline
\textbf{index} & \textbf{MP ID} & \textbf{set} & \textbf{no. configurations} & \textbf{no. Hf} & \textbf{no. O} \\ \hline
1              & --             & Hf\textsubscript{2} gas          & 63                          & 2               & 0              \\
2              & 103            & Hf\textsubscript{2} EOS 1D       & 202                         & 2               & 0              \\
3              & 103            & Hf\textsubscript{2} EOS 3D       & 9377                        & 2               & 0              \\
4              & 103            & Hf\textsubscript{2} EOS 6D       & 17205                       & 2               & 0              \\
5              & 100            & Hf EOS 1D        & 201                         & 1               & 0              \\
6              & 100            & Hf 1D primitive  & 201                         & 1               & 0              \\
7              & 100            & bulk Hf MC       & 306                         & 128             & 0              \\
8              & 103            & bulk Hf MC       & 50                          & 128             & 0              \\
9              & --             & bulk Hf MC       & 498                         & 128             & 0              \\
10             & 607540         & O\textsubscript{2} EOS 6D        & 19223                       & 0               & 2              \\
11             & --             & O\textsubscript{2} gas           & 204                         & 0               & 2              \\
12             & --             & O EOS 6D         & 19689                       & 0               & 1              \\ \hline
\end{tabular}
\label{tab:testdata}
\end{table}

Note that in Figure \ref{fig:prediction}, the bulk Hf test data corresponds to sets 1-3,  the Hf\textsubscript{2} test data to sets 4-7, the Hf test data to sets 8-9, the O\textsubscript{2} test data to sets 10-11, and the O test data to set 12, as listed in Table \ref{tab:testdata}.

\subsection{Additional experiments on hafnium data}
\label{app:diversity}

\begin{figure}[hbt!]
    \centering
    \includegraphics[width=0.85\textwidth]{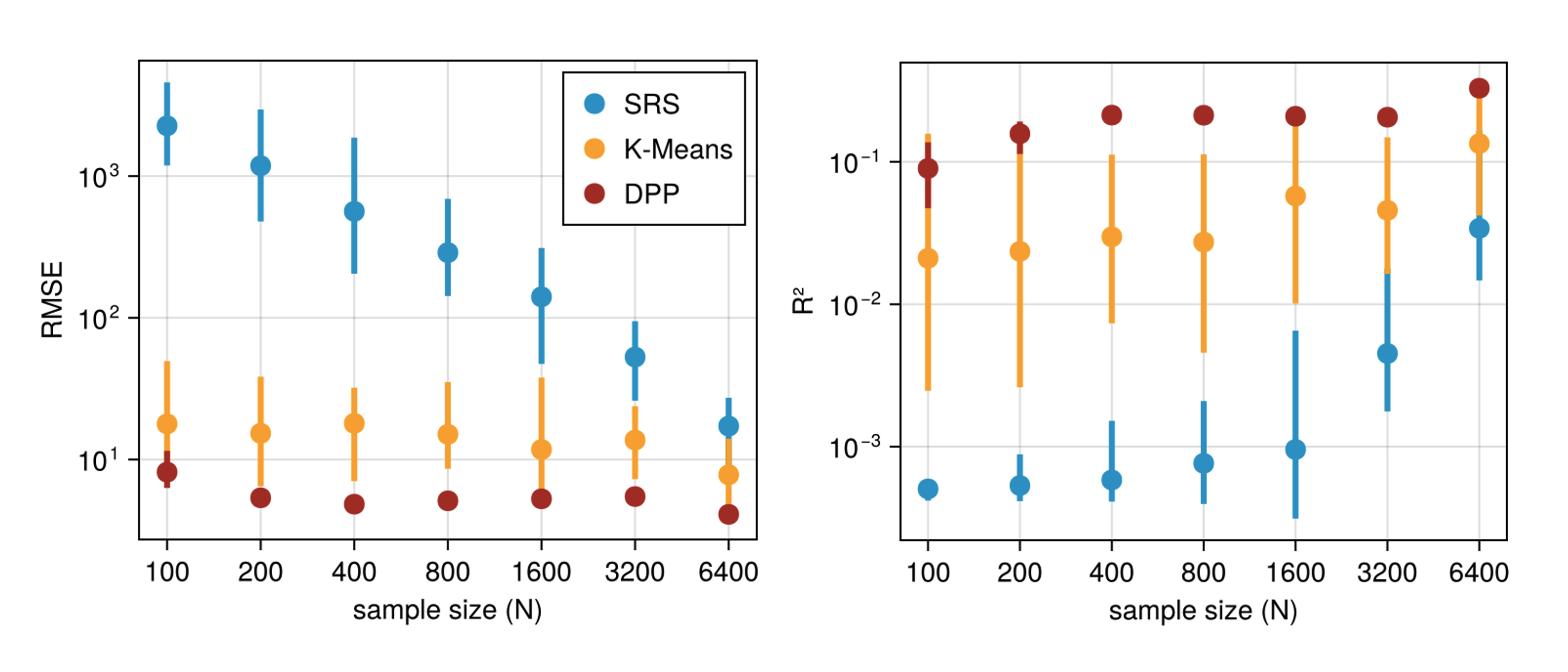}
    \caption{RMSE and $R^2$ values of energy predictions on the validation set by the hafnium MLIP for training set size $N$. The center point denotes the median value and rangebars denote the interquartile range (25th to 75th percentile) over 100 trials.}
    \label{fig:train_accuracy_hf}
\end{figure}

\begin{figure}[hbt!]
    \centering
    \includegraphics[width=0.72\textwidth]{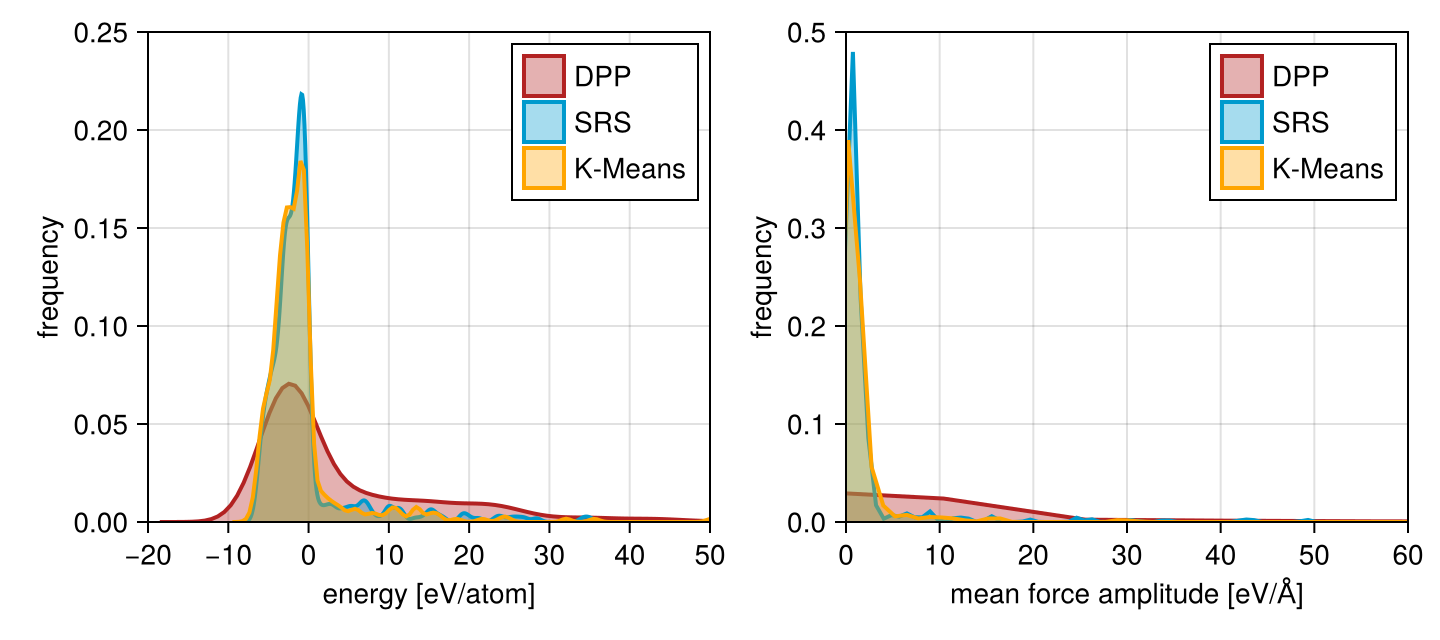}
    \caption{Distribution of energies and force amplitudes of 400 configurations selected by each algorithm.}
    \label{fig:ef_rep_hf}
\end{figure}

In another study, we consider single-species hafnium (Hf) data to evaluate the performance of DPP, $k$-means ($k=5$), and SRS to draw variable-sized training datasets. We take the candidate training data and validation data from a 27,249-element set of 1-atom and 2-atom Hf configurations, generated from simulation sets 1-6 detailed in Table \ref{tab:testdata}. The training sets, subselected from a 70\% split of the data, are used to learn a 5-body 6-degree ACE potential of hafnium, which have a descriptor dimension of $d=35$. Figure \ref{fig:train_accuracy_hf} shows the statistics of RMSE and $R^2$ values of the energy predictions on the validation set by MLIPs trained on each subset. Figure \ref{fig:ef_rep_hf} shows the distribution of per-atom energies and force magnitudes associated with a 400-sample set drawn by each of the three methods. While the results using the Hf data largely mirror the trends observed with the Hf-O data in Section \ref{sec:experiment}, the improvements of the DPP-based approach in terms of both prediction accuracy and diversity of configurations are more pronounced in this study. Further experiments can be conducted to assess the dependency of the data subselection algorithms on the choice of descriptor, in particular the dimension and expressivity of the descriptor for a given system of study.

\begin{figure}[hbt!]
    \centering
    \includegraphics[width=0.6\textwidth]{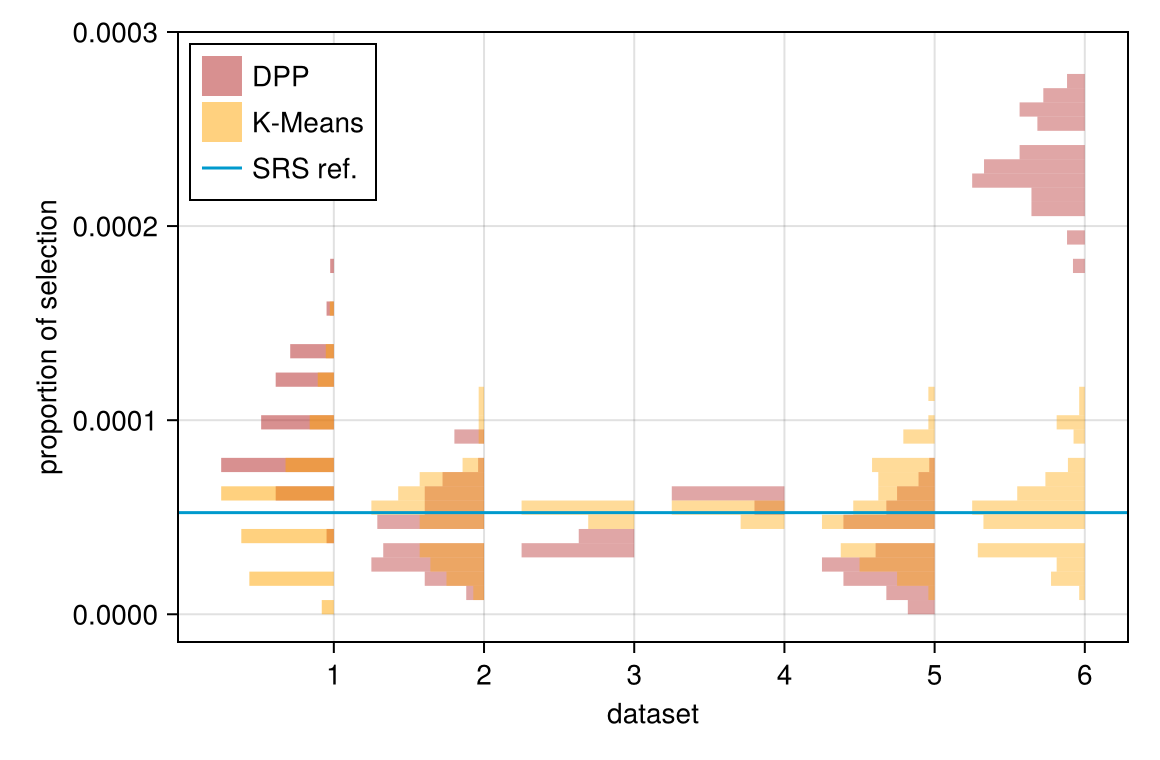}
    \caption{Representation from simulation datasets 1-6 (Table \ref{tab:testdata}) in the 800-element subsets selected by DPPs and $k$-means, with the uniform rate from SRS as reference.}
    \label{fig:ds_rep}
\end{figure}

In addition, we assess diversity of subsets selected by DPP and $k$-means in terms of the rate at which the algorithm samples each of the 6 simulation datasets composing the candidate training data (the first six rows of Table \ref{tab:testdata}). Over 100 trials, subsets of 800 elements are drawn from the data pool of $M_{\text{tot}} = 19,090$ elements and the number of instances of each simulation dataset is counted and normalized by the batch size and total number of elements per dataset, such that the normalized rate can be compared to the uniform sampling rate of $1/M_{\text{tot}}$ from SRS. Figure \ref{fig:ds_rep} shows that the DPP leads to differential rates of sampling between datasets, with Sets 1 and 6 favored relative to the SRS baseline. This indicates that these two simulation sets, while relatively small in number compared to other sets, are important to include in the DPP subset to represent diversity in the data pool. Therefore, DPPs can also be utilized as a diagnostic for categories of data which have proportionally greater influence in training, which can inform the collection of additional data. In contrast, the $k$-means clustering method does not lead to substantially different sampling rates compared to SRS. 

\subsection{Determinantal point processes}
\label{app:dpps}

Consider a set of $M$ discrete elements represented by their indices $\mathcal{Y} = \{1,...,M\}$.  A determinantal point process (DPP) is a probability measure placed over all $2^M$ subsets of $\mathcal{Y}$, where probabilities are determined by the kernel matrix $K \in \mathbb{R}^{M \times M}$ associated with the process. In practice, the kernel matrix is constructed from evaluations of a positive semidefinite (PSD) kernel function $\kappa: \mathcal{Y} \times \mathcal{Y} \to \mathbb{R} $ between each pair of elements in the set, where $K_{ij} = \kappa(Y_i, Y_j)$ for $Y_i, Y_j \in \mathcal{Y}$. If the kernel matrix satisfies conditions for the existence of the $L$ formulation (namely, that $P(\emptyset) \neq 0$ and $K$ has no eigenvalue equal to 1 \citep{Kulesza2012}), then the $L$ ensemble  corresponding to $K$ is given by: 

\begin{equation}
    \label{eq:ell}
    L = K (I - K)^{-1}
\end{equation}

The DPP is then defined equivalently by the following, for random subsets $\textcolor{blue}{Y} \subseteq \mathcal{Y}$ and a fixed subset $A \subseteq \mathcal{Y}$ \citep{Edelman}:

\textit{PDF definition.} The probability density function of the DPP is given by: 

\begin{equation}
    \mathcal{P}(\textcolor{blue}{Y} = A) = \frac{\det(L_A)}{\sum_{A' \subseteq \mathcal{Y}} \det(L_{A'})}
\end{equation}
where $L_A = [ L_{ij} ]_{i,j \in A}$ denotes the matrix restricted to entries indexed by the elements of A. The PDF definition is also referred to as the ``$L$ formulation'' of the DPP \citep{Edelman}. 

\textit{CCDF definition.} The complementary cumulative density function of the DPP is given by: 

\begin{equation}
    \mathcal{P}(\textcolor{blue}{Y} \supseteq A) = \det(K_{A})
\end{equation}
In other words, the probability that $A$ is a subset of the randomly drawn set $\textcolor{blue}{Y}$ is given by the determinant of the kernel matrix restricted to entries indexed by $A$. A special case of the CCDF is the marginal probability of each element of the set, which is given by the diagonal of the $K$ matrix: 

\begin{equation}
    \mathcal{P}(i \in \textcolor{blue}{Y}) = K_{ii}
\end{equation}
The marginal probability is also referred to in literature as the ``inclusion probability'' \citep{Kulesza2012}. The CCDF definition is also referred to as the ``$K$ formulation'' of the DPP \citep{Edelman}. 

\textit{CDF definition.} The cumulative density function of the DPP is given by: 

\begin{equation}
    \mathcal{P}(\textcolor{blue}{Y} \subseteq A) = \det(I - K)_{\bar{A}}
\end{equation}

where $\bar{A}$ denotes the complement, $\bar{A} = \mathcal{Y} \setminus A$. 
\bigskip

\textbf{Mixture of elementary DPPs.} 
An important property of a DPP is that it can be represented as the mixture of elementary DPPs. Also referred to as \textit{projection DPPs}, an elementary DPP has a kernel matrix which is a projection matrix of rank $r \leq M$, e. g. $K^\textup{T} K = K$ and $K = VV^\textup{T}$ for a set of $r$ orthonormal vectors $V$ \citep{Edelman}. Elementary DPPs then have the property that: 

\begin{equation}
  \mathcal{P}^{V_r}(A) =
    \begin{cases}
      \det(K_A) & \text{if $|A| = r$}\\
      0 & \text{otherwise}
    \end{cases}       
\end{equation}

Therefore, only $\binom{M}{r}$ subsets of size exactly $r$ have non-zero probability \citep{Edelman}. The PDF of the DPP can be represented as the mixture of elementary DPPs, using the eigendecomposition of the $L$ matrix $L = \sum_{i=1}^M \lambda_i \mathbf{v}_i \mathbf{v}_i^{\textup{T}}$:

\begin{equation}
    \mathcal{P}(\textcolor{blue}{Y} = A) = \frac{1}{\det(I+L)} \sum_{J \subseteq 1:M} \mathcal{P}^{V_J} \prod_{i \in J} \lambda_i
\end{equation}

In particular, it can be shown that the normalizing constant of the density can be derived as in \cite{Kulesza2011}: 

\begin{equation}
    \sum_{A' \subseteq \mathcal{Y}} \det(L_{A'}) = \det(I + L) = \prod_{i=1}^M (\lambda_i + 1)
\end{equation}

The mixture representation of DPPs lends it to a computationally tractable sampling algorithm. In particular, the DPP can be sampled by drawing samples from each of the elementary DPPs with probability $\frac{\prod_{i \in J} \lambda_i}{\prod_{i=1}^M (\lambda_i + 1)}$.
\bigskip

\textbf{Fixed-size determinantal point processes.} A $k$-DPP is a DPP which produces samples of fixed size $k \leq M$ (not to be misconstrued with the number of clusters in $k$-means). Unlike elementary DPPs, which are restricted to represent specific probability measures associated with a projection kernel matrix, $k$-DPPs can represent a more flexible range of probability measures over the subsets. As one example, a $k$-DPP can be defined to assign a uniform distribution over subsets, whereas a singular elementary DPP cannot \citep{Kulesza2012}. Therefore, elementary DPPs can be considered a subclass of $k$-DPPs. 

A $k$-DPP can be understood as a special form of conditional DPP, with the following PDF: 

\begin{equation}
    \mathcal{P}(\textcolor{blue}{Y} = A | \  |\textcolor{blue}{Y}| = k) = \frac{\det(L_A)}{\sum_{|A'| = k} \det(L_{A'})}
\end{equation}

where the normalizing constant is a sum over all subsets $A' \in \mathcal{Y}$ with restricted cardinality $|A'| = k$. It can be shown that the $k$-DPP can also be expressed as a mixture of elementary DPPs: 

\begin{equation}
    \mathcal{P}(\textcolor{blue}{Y} = A | \  |\textcolor{blue}{Y}| = k) = \frac{1}{e^M_k} \sum_{|J| = k} \mathcal{P}^{V_J} \prod_{i \in J} \lambda_i
\end{equation}

The normalizing constant of this distribution differs from that of the standard DPP. One can show it is derived as:

\begin{equation}
    \begin{aligned}
        \sum_{|A'| = k} \det(L_{A'}) & = \det(I+L) \sum_{|A'| = k} \mathcal{P}(\textcolor{blue}{Y} = A') \\
        & = \sum_{|J| = k} \prod_{i \in J} \lambda_i
    \end{aligned}
\end{equation}

The derivation uses the property that sets drawn from the elementary DPP have cardinality $|J|=k$ with probability 1, such that the expression reduces to the sum of products of eigenvalues indexed by elements in each $J$ subset. One can recognize this term to be the $k$th elementary symmetric polynomial \citep{Kulesza2012}: 

\begin{equation}
    e^M_k = e_k(\lambda_1, ..., \lambda_M) = \sum_{J \subseteq 1:M \atop |J| = k} \prod_{i \in J} \lambda_i
\end{equation}

The marginal probability of elements, now considering fixed sizes to the subsets drawn, is proportional to the eigenvalues of $L$ scaled by a ratio of the elementary symmetric polynomials: 

\begin{equation}
    \label{eq:marginalprob}
    \mathcal{P}(i \in \textcolor{blue}{Y} | \  |\textcolor{blue}{Y}| = k) = \lambda_M \frac{e^{M-1}_{k-1}}{e^M_k}
\end{equation}

\bigskip

\textbf{Sampling algorithm for $k$-DPPs}. Algorithm \ref{alg:sampling} for sampling from $k$-DPPs is reproduced from \cite{Kulesza2011, Kulesza2012}. The algorithm is composed of two loops: Loop 1 samples eigenvectors of $L$ to form a subspace from which to draw the samples. While the eigenvectors are sampled with probability $\frac{\lambda_n}{\lambda_n+1}$ for standard DPPs, the probability becomes $\lambda_n \frac{e^{n-1}_{l-1}}{e^n_l}$ for $k$-DPPs. Moreover, sampling is performed until strictly $k$ eigenvectors are obtained. Loop 2 iteratively samples elements from the eigenvectors, orthonormalizing the basis after each sample is drawn. While the cardinality of the basis $V$ varies for standard DPPs, it is kept fixed at $k$ for $k$-DPPs. 

Computationally, the sampling of regular DPPs and $k$-DPPs differ primarily in the computation of the probabilities in Loop 1. For $k$-DPPs, the normalization constant of the probability density involves calculating the ratio of elementary symmetric polynomials, the cost of which scales exponentially. For instance, calculation of $e^N_k$ takes on the order of $\mathcal{O}(k \binom{N}{k})$ operations due to the combinatorial problem. To address this cost, \cite{Kulesza2011, Kulesza2012} recommend implementing a recursive algorithm to construct all symmetric elementary polynomials at once, using the recurrence relation: 

\begin{equation}
    e^N_k = e^{N-1}_k + \lambda_N e^{N-1}_{k-1}
\end{equation}

The recursive algorithm has polynomial cost at $\mathcal{O}( N k )$, significantly reducing the computational overhead of the sampling algorithm.

\begin{algorithm}
\caption{Sampling from a k-DPP}\label{alg:sampling}
\begin{algorithmic}
\Require $0 < k \leq N$, eigendecomposition $\{(\mathbf{v}_n, \lambda_n)\}_{n=1}^N$ of $L$ from Equation \ref{eq:ell}
\\
\State \textit{Loop 1: sample eigenvectors to form subspace}
\State $J \gets \emptyset$
\State $l \gets k$
\For{n = N, ..., 2, 1}
    \If{$l = 0$}
        \State \textbf{break}
    \EndIf
    \If{$u \sim U[0,1] < \lambda_n \frac{e^{n-1}_{l-1}}{e^n_l}$}
        \State $J \gets J \cup \{ n \}$
        \State $l \gets l - 1$
    \EndIf
\EndFor
\\
\State \textit{Loop 2: draw samples from orthonormalized subspace}
\State $V \gets \{\mathbf{v}_n \}_{n \in J}$
\State $Y \gets \emptyset$
\While{$|V| > 0$}
    \State Select $i$ from $\mathcal{Y}$ with $Pr(i) - \frac{1}{|V|} \sum_{\mathbf{v} \in V} (\mathbf{v}^\textup{T} \mathbf{e}_i)^2$
    \State $Y \gets Y \cup i$
    \State $V \gets V_\perp$ (orthonormal basis for subspace of $V$ orthogonal to $\mathbf{e}_i$)
\EndWhile
\\
\\
\textbf{Output:} $Y$
\end{algorithmic}
\end{algorithm}


\subsection{Implementation}

Experiments were conducted in Julia using the packages \href{https://github.com/cesmix-mit/PotentialLearning.jl}{PotentialLearning.jl} and \href{https://github.com/cesmix-mit/InteratomicPotentials.jl}{InteratomicPotentials.jl}, developed Dallas Foster, Emmanuel Lujan, Spencer Wyant, and JZ; \href{https://github.com/dahtah/Determinantal.jl}{Determinantal.jl}, developed by Simon Barthelm\'e; \href{https://github.com/Muxas/Maxvol.jl}{Maxvol.jl}, developed by Aleksandr Mikhalev; \href{https://github.com/ACEsuit/ACE.jl}{ACE.jl}, developed by Christoph Ortner et al.; and \href{https://github.com/JuliaStats/Clustering.jl}{Clustering.jl} for the $k$-means algorithm.